\documentstyle[12pt]{article}
\newcommand{\plabel}{\label}
\begin{document}
\vskip .9cm
{\large\bf
 \centerline{$SU(2)$--invariant reduction of the 3+1 dimensional}
 \centerline{ Ashtekar's gravity} }
\vskip .7cm

\centerline{Sergei Alexandrov{${}^\dag $}, Ignati Grigentch{${}^\ddag $}
 and Dmitri Vassilevich{${}^\dag$}  }

\vskip .7cm

${}^\dag${\it  Department of Theoretical Physics, St. Petersburg
        University 198904,

St. Petersburg, Russia }
\vskip .3cm
  ${}^\ddag${\it Department of Physics, Old Dominion University, Norfolk,
  VA 23529 
  \centerline{and} 
  
  Jefferson Lab, 12000 Jefferson Avenue, Newport News, VA 23606, USA }

\vspace{15mm}

\centerline{\bf Abstract}

We consider a space--time with spatial sections isomorphic
to the group manifold of $SU(2)$. Triad and connection fluctuations
are assumed to be $SU(2)$-invariant. Thus, they form a finite
dimensional phase space. We perform non--perturbative path
integral quantization of the model. Contarary to previous 
claims the path integral measure appeared to be non--singular
near configurations admitting additional Killing vectors.
In this model we are able to calculate the generating 
functional of Green functions of the reduced phase space 
variables exactly.

\newpage
\setcounter{page}{2}

\section{Introduction}
Ashtekar's variables \cite{Ash} simplify considerably the
algebraic structure of constraints in general relativity.
These variables give rise to even simpler models \cite{mod},
which can be solved completely. Such models are useful for a
non--perturbative study of quantum gravitational effects.

In the present paper we start with a $3+1$-dimensional Ashtekar's
gravity. We assume that spatial sections are topologically
$S^3$, which is the group manifold of $SU(2)$. Next we reduce the phase
space to $SU(2)$ invariant fields. This gives a finite dimensional
(complex) phase space. By solving constraints and fixing gauge
freedom we arrive at a four dimensional (real) reduced phase
space\footnote{In our model the phase space of Ashtekar's gravity
undergoes two subsequent reductions. We hope, it is clear from
the context, which one is meant in any particular case.}.
We also construct the path integral for this model and calculate
Green functions at all orders of the perturbation theory. The corresponding
quantum effective action coincides with the classical one. Our
technique is similar to that used in the 2D dilatonic gravity \cite{KLV}.

Our primary aim is to study the behavior of the path integral measure
near the points in the phase space admitting additional Killing vectors.
Mottola \cite{Mot} argued that due to the presence of zero modes
in the Faddeev--Popov determinant on a background with Killing
vectors the path integral measure become zero, and hence
symmetric configurations (as e.g. de Sitter space) do not
contribute to the path integral. Our model appeared to be useful
for a non--perturbative study of this effect. Among $SU(2)$-invariant
configurations it contains also $SU(2)\times U(1)$ and
$SU(2)\times SU(2)$-invariant configurations. We found that
the Faddeev--Popov determinant has zeros at symmetric
configurations. However, these zeros are cancelled by the
contribution of the delta functions of the constraints. The
resulting path integral measure is regular.
\section{The reduced action}
We begin with a complex gravitational action in $3+1$ dimensions,
\begin{equation}
  S=\int d^4x\left (i{E_a}^i\partial_t{A_i}^a-N^aG_a-N^iG_i-NG_0\right ),
\plabel{act}
\end{equation}
where, as usual, the densitized triad ${E_a}^i$ and connection ${A_i}^a$ are
the canonical variables;\ $G_a,G_i$ and $G_0$ are the Gauss law, the vector
and the scalar constraints respectively:

\begin{eqnarray}
G_a&=&{\cal D}_i{E_a}^i=0,\\
G_i&=&{F_{ij}^a}{E_a}^j=0,\\
G_0&=&\varepsilon^{abc}{E_a}^i{E_b}^j{F_{ij}^c}+{\Lambda \over 3}
\varepsilon^{abc}\varepsilon_{ijk}{E_a}^i{E_b}^j{E_c}^k=0,
\end{eqnarray}
where $\Lambda$ is the cosmological constant; $N^a, N^i$ and $N$ are the 
Lagrange multipliers, ${\cal D}_i$ is the covariant derivative with respect 
to the connection ${A_i}^a$ , $F_{ij}^a$ is the field strength.

We assume that the spatial sections are topologically equivalent to $S^3$.
We restrict ourselves to the $SU(2)$--invariant canonical variables ${A_a}^i$
and ${E_i}^a$. They can be presented in the following form:
\begin{equation}
 {A_a}^i={f_b}^a\left (t\right ){e_i}^b \plabel{Aai}
\end{equation}
\begin{equation}
{E_a}^i={h_a}^b\left (t\right ){e_b}^ie,\plabel{Eai}
\end{equation}
where ${e_i}^a$ is an invariant triad field on ``round'' $S^3$,
$e=\det \left ({e_i}^a\right )$
and $f,h$ are
 $3\times 3$ (complex) matrices depending only on time. Here we used the
fact that $S^3$ is the group manifold of $SU(2)$. In general, the
canonical connection $A^{[c]}$ should be subtracted from the r.h.s.
of (\ref{Aai}). In the canonical coordinates on $SU(2)$ (see bellow)
$A^{[c]}$ is zero, and $A_i^a$ behaves like a tensor with respect to
$SU(2)$ transformations. For any given group ${\cal G}$ the ${\cal
G}$-invariant tensor fields are those that have constant components in the
canonical tagential basis.  The case of unit matrices $f$ and $h$
corresponds to the $SU(2)\times SU(2)$-invariant configuration isometric to
the ``round'' $S^3$ with the metric of the maximum symmetry.

Group elements of $SU(2)$ can be considered as canonical coordinates on $S^3$. 
 We can take $g=\exp (x^j \frac 1{2i} \tau_j)$, where $\tau_j$ are the Pauli
 matrices. 
The triad one-form can be calculated from the relation
$e=g^{-1}dg$. In the vicinity of the unit element we have
\begin{equation}
e_j^b(0)=\delta_j^b,\quad \partial_j e_k^b(0)= -\frac 12
\varepsilon_{jkb}. \plabel{triad}
\end{equation}
Regardless of the type of
the indices the Levi--Civita symbol $\varepsilon$ is $\pm 1$.

By substituting the ansatz (\ref{Aai}), (\ref{Eai}) into the
action (\ref{act}) we obtain:
\begin{equation}
  S=\int dt\left (i{h_a}^b\partial_t{f_b}^a-
\eta^a{\cal C}_a-n^a{\cal H}_a-n{\cal H}_0\right ),
\plabel{act1}
\end{equation}
where only $SU(2)$ invariant components of the Lagrange multipliers
survive:
\begin{equation}
N^a=\eta^a(t), \quad N^i=e^i_a n^a(t), \quad eN=n(t).
\plabel{Lmul}
\end{equation}
In (\ref{act1}) we discarded an overall constant factor equal to
the volume of $SU(2)$.

The Poisson brackets of the canonical variables become:
\begin{equation}
\left \{ {h_a}^b\ ,\ {f_d}^c \right \}=-i{\delta_a}^c{\delta_d}^b.
\end{equation}

The following useful identity holds on the group manifold of $SU(2)$:
\begin{equation}
e_a^k\partial_ke_b^j-e_b^k\partial_ke_a^j=
[e_a,e_b]^j=\varepsilon_{abc}e_c^j,
\end{equation}
which enables us to express the constraints in terms of the invariant
variables
\begin{eqnarray}
{\cal C}_a&=&\varepsilon_{abc}{f_d}^b{h_c}^d
\plabel{Ca} \\
{\cal H}_a&=&\varepsilon_{abc}{f_b}^d{h_d}^c+
\varepsilon_{ebc}{f_d}^b{h_c}^d{f_a}^e
\plabel{Ha} \\
{\cal H}_0&=&
-\varepsilon^{abc}\varepsilon_{def}{h_a}^{d}{h_b}^{e}{f_{f}}^c
+\varepsilon^{abc}\varepsilon_{fgc}{f_d}^{f}{h_a}^d{f_e}^{g}{h_b}^e+
\nonumber \\ \ &\ &+{\Lambda \over
3}\varepsilon^{abc}\varepsilon_{def}{h_a}^{d}{h_b}^{e}{h_c}^{f}.
\plabel{H}
\end{eqnarray}

For the sake of symmetry it is more convenient, however, to consider 
a combination of the vector and Gauss law constraints (see \cite{Peldan}) 
instead of the vector constraint itself.  The modified vector constraint reads

\begin{equation}
{\tilde G}_i\ =G_i-{A_i}^aG_a.\plabel{tilG}
\end{equation}
In terms of the invariant variables this constraint becomes
\begin{equation}
\tilde {\cal H}_a=-\varepsilon^{abc}{h_d}^b{f_c}^d \plabel{tilH}
\end{equation}
Instead of the original action (\ref{act1}) we can use
the following action
\begin{equation}
  S=\int dt\left (i{h_a}^b\partial_t{f_b}^a-
\tilde \eta^a{\cal C}_a- n^a\tilde {\cal H}_a-n{\cal H}_0\right ),
\plabel{act2}
\end{equation}
which is obtained from (\ref{act1}) by a shift of the Lagrange multiplier $\eta^a$.

Now we are to find gauge transformations of the canonical variables. In
general, an infinitesimal gauge transformation generated by a constraint $G$
of a variable $Z$ is

\begin{equation}
\delta Z=\left \{G\xi\ ,\ Z\right \},
\end{equation}
where $\xi$ is the infinitesimal parameter of the transformation.

In our case the constraints (\ref{Ca}), (\ref{H}) and (\ref{tilH}) lead
to the following transformations:

Gauss law constraint:
\begin{eqnarray}
\delta {h_a}^d&=&i\varepsilon_{abc}{h_b}^d{\xi}^c
\nonumber \\
\delta {f_d}^a&=&i\varepsilon^{abc}{f_d}^b{\xi}^c ,
\plabel{gautr}
\end{eqnarray}

modified vector constraint:
\begin{eqnarray}
\delta {h_d}^a&=&i\varepsilon^{abc}{h_d}^b\tilde {\xi}^c
\nonumber \\
\delta {f_a}^d&=&i\varepsilon_{abc}{f_b}^d\tilde {\xi}^c ,
\plabel{vectr}
\end{eqnarray}

scalar constraint:
\begin{eqnarray}
\delta {h_d}^a&=&2i\xi {h_d}^a{h_b}^c{f_c}^b - 2i\xi
{h_b}^a{f_c}^b{h_{d}^c -
i\xi \varepsilon_{dgh}\varepsilon^{abc}{h_g}}^b{h_{h}}^c
\nonumber \\
\delta {f_d}^a&=&2i\xi
\varepsilon^{abc}\varepsilon_{dfh}{h_b}^{f}{f_{h}}^c -
2i\xi {f_d}^a{f_c}^b{h_b}^c \nonumber \\
\ &\ &+ 2i\xi {f_b}^a{h_c}^b{f_d}^c - i\Lambda \xi
\varepsilon^{abc}\varepsilon_{def}{h_b}^{e}{h_c}^{f}.
\plabel{sctr}
\end{eqnarray}

The non-zero Poisson brackets of the constraints are:
\begin{eqnarray}
\{ {\cal C}_a,{\cal C}_b \} &=& i\varepsilon_{abc}{\cal C}_c
\nonumber \\
\{ \tilde {\cal H}_a,\tilde {\cal H}_b \} &=&
i\varepsilon_{abc}\tilde {\cal H}_c \plabel{alg}
\end{eqnarray}
Thus the constraint algebra splits into a direct sum of two
copies of $so(3,C)$ and a one-dimensional abelian algebra.

\section{Reduced phase space quantization}
To construct a reduced phase space one should solve the constraints
(\ref{Ca}), (\ref{H}), (\ref{tilH}) and fix the corresponding gauge freedom.

The symmetry between the Gauss law (\ref{Ca}) and the modified vector
constraint (\ref{tilH}) makes it natural to eliminate the corresponding
gauge freedom (\ref{gautr}) and (\ref{vectr}) simultaneously. One can
integrate these infinitesimal transformations to finite gauge transformations. 
Then two of them rotate both the upper and the lower indices of $h$ by $SO(3,C)$ matrices.
One can see that by means of combined action of these two transformations the 
matrix $h$ can be diagonalized:

\begin{equation}
 {h_a}^b={\rm diag}(h_1, h_2, h_3). \plabel{hdiag}
\end{equation}

Now we are to use this condition to solve both the Gauss law and the modified
vector constraint. There are several different cases to be considered:

\begin{enumerate}
\item
 Non-degenerate fluctuation of the triad.\\
 All three elements of $h$ have different absolute values:
 \begin{equation}
  |h_1|\neq |h_2|\neq |h_3|.
 \end{equation}
 In this case the solution of both (\ref{Ca}) and (\ref{tilH})
is the diagonal fluctuation of the connection:
 \begin{equation}
   {f_a}^b={\rm diag}(f_1, f_2, f_3). \plabel{fdiag}
 \end{equation}
\item
 Once degenerate fluctuation of the triad.\\
 Two of the three elements of $h$ have equal absolute values:
 \begin{equation}
  |h_1|= |h_2|\neq |h_3|
 \end{equation}
If $h_1=h_2,$ the solution for $f$ can be written as
 \begin{equation}
 \pmatrix {f_{11}& f & 0\cr f & f_{22} & 0\cr 0 & 0 & f_{33}\cr} .
 \plabel{1deg}
 \end{equation}

 Note that in the case of once degenerate $h$ only five of the six
 parameters of the gauge transformations (\ref{gautr})-(\ref{vectr})
 are fixed by choosing the diagonal form of the triad fluctuation. For
 example, if $h_1=h_2$ the transformation parametrized by
 \begin{equation}
 \xi^c=\tilde \xi^c \propto \delta_3^c
 \end{equation}
 is not involved in the diagonalization of $h$ and hence can be used to
 diagonalize (\ref{1deg}). This fixes the gauge freedom completely unless
 the matrix
 $\pmatrix{f_{11}&f\cr f& f_{22}}$ has equal eigenvalues. If the
 latter is the case this part of the gauge freedom can not be eliminated by any
 condition on the phase variables $f$ and $h.$ A similar situation for ADM
 gravity was considered in \cite{DIJMPA} and for Ashtekar's gravity on a de
 Sitter background in \cite{GV}. The proper way to fix the remaining
 freedom is to impose a condition on a Lagrange multiplier.

\item
 Twice degenerate fluctuation of the triad.\\
 All three components of $h$ have equal absolute values:
 \begin{equation}
 |h_1|=|h_2|=|h_3|.
 \end{equation}
 In this case the general solution ${f_a}^b$ of the constraints
(\ref{Ca}) and (\ref{tilH}) has six independent components. For
  example, for $h_1=h_2=h_3,$ any symmetric fluctuation of the
 connection solves both the Gauss law and the modified vector constraint.

Just like in the previous case the part of gauge freedom not fixed by
the condition $(19)$ can be utilized to diagonalize $f.$ Three different
off-diagonal elements of $f$ can be set to zero by the three parameter
gauge transformation of the form:
\begin{equation}
 \xi^c=\tilde \xi^c.
\end{equation}
And again this works only for a nondegenerate fluctuation of the
connection, i.e. when eigenvalues of $f_a^b$ are all different.
Otherwise an additional condition on a Lagrange multiplier should be
imposed.  \end{enumerate}

The last two cases of degenerate triad fluctuations correspond to additional
invariance of the field configuration and existence of additional Killing
vectors. For example, the twice degenerate case corresponds to additional
$SO(3)$ invariance. The necessity to exclude some of the Lagrange
multipliers is natural in the context of Hamiltonian theory
because some of the constraints become linearly dependent. Imposing
a new gauge condition on a submanifold in a phase space looks awkward from the
point of view of an ordinary gauge theory. However, this is just a
manifestation of the Gribov problem: there is no global gauge
fixing in a non-abelian gauge theory. This is especially clear
in our case since gauge orbits in the phase space have different dimensions.

Before considering the remaining constraint (\ref{H}), let us
study the reality conditions. In our case they read:
\begin{equation}
{\rm Im\ }h_a^b=0, \quad {\rm Re\ }f_a^b=-\frac 12 \Gamma_a^b (h)
\plabel{recon}
\end{equation}
where $\Gamma_a^b=e_a^i\varepsilon^{bcd}\omega^{cd}_i$, \ 
$\omega_i$ is the spin connection compatible with the rescaled
inverse triad ${h_a}^be_b^i$.
For the gauge (\ref{hdiag}) the spin--connection $\Gamma$ is
diagonal,
\begin{equation}
\Gamma (h) ={\rm diag} (\Gamma_1,\Gamma_2,\Gamma_3),
\plabel{Ga1}
\end{equation}
with
\begin{eqnarray}
\Gamma_1&=&\frac {h_2h_3}{h_1^2} -\frac {h_2}{h_3} -\frac {h_3}{h_2}
\nonumber \\
\Gamma_2&=&\frac {h_1h_3}{h_2^2} -\frac {h_1}{h_3} -\frac {h_3}{h_1}
\nonumber \\
\Gamma_3&=&\frac {h_2h_1}{h_3^2} -\frac {h_2}{h_1} -\frac {h_1}{h_2}.
\plabel{Ga2}
\end{eqnarray}
We see, that the reality conditions (\ref{recon}) are consistent
with the solutions (\ref{fdiag}) of the constraint equations.
Note, that the imaginary part of ${\cal H}_0$ vanishes.

We use reality in the "triad" and not in the "metric"
form\footnote{Discussion on different types of reality
conditions see e.g. in \cite{YoSh}.}. It is essential for
the path integral quantization because we wish to restrict
the integration variables rather than their composites.
Imposing reality conditions restricts the gauge group to
its real form. Thus, it is important to fix all complex
gauge transformations but this should be done in a way
consistent with the reality conditions.

It is convenient to fix the remaining gauge freedom (\ref{sctr})
by the condition
\begin{equation}
h_1=1 .\plabel{hamfix}
\end{equation}
The real part of ${\cal H}_0$ gives
\begin{eqnarray}
\phi_1=\frac 1{h_2\phi_2+h_3\phi_3} \times \nonumber \\
\left (
\frac 14 \left ( \left ( \frac {h_2}{h_3} \right )^2 +
\left ( \frac {h_3}{h_2} \right )^2 + (h_2h_3)^2 \right )
-\frac 12 (h_2^2+h_3^2+1)+\Lambda h_2h_3
-h_2h_3\phi_2\phi_3 \right ), \plabel{f}
\end{eqnarray}
where $\phi_a={\rm Im}\ f_a$.

Now we are able to calculate the path integral measure
\cite{SF}
\begin{equation}
d\mu =\prod_{i,j} (df_i^j\ dh_i^j) \det |\{ G_A,\chi^B \} |
\prod_A\delta (\chi^A )\delta (G_A), \plabel{meas}
\end{equation}
where $G_A$ denotes all constraints, and $\chi^B$ stand for
corresponding gauge conditions. Let us subdivide the phase
space variables into two classes: $\{ f,h\} =
\{ f_2,f_3,h_2,h_3,f_A,h^A\}$ where the first four variables
are independent, while $f_A$ and $h^A$ are fixed by means
of constraints or gauge conditions. Then,
\begin{equation}
\{ G_A,\chi^B\} =
\{ G_A,h^B\} = \frac {\partial G_A}{\partial f_B}.
\plabel{GAf}
\end{equation}
Being combined with $\delta (G_A)$ the determinant of
(\ref{GAf}) gives $\delta (f_A-f_A^{(0)})$, where
$f_A^{(0)}$ are solutions of the constraints expressed
in terms of independent variables. Hence the measure
(\ref{meas}) becomes
\begin{equation}
d\mu = df_2\ df_3\ dh_2\ dh_3. \plabel{mea1}
\end{equation}
In the integrand all phase space variables should be expressed
in terms of $f_2,\ f_3, h_2,\ h_3$. Since the shift of $f$ by
$\frac 12 \Gamma (h)$ is a canonical transformation, it does not
change the measure. This means that the real integration measure
will be just
\begin{equation}
d\mu_R=d\phi_2\ d\phi_3\ dh_2\ dh_3, \plabel{Rme}
\end{equation}
where we now integrate over real variables.

The generating functional of Green functions is
\begin{eqnarray}
Z(J,j)=\int d\mu_R \exp i\int dt (&-&h_2 \partial_t \phi_2 -
h_3 \partial_t \phi_3 +\frac i2 \Gamma_2\partial_th_2+
\frac i2 \Gamma_3\partial_th_3 \nonumber \\
&+&h_2J^2+h_3J^3+\phi_2j^2+\phi_3j^3 ).
\plabel{Z}
\end{eqnarray}
Integration over $\phi_2$ and $\phi_3$ gives delta functions
\begin{equation}
\delta (\partial_th_2+j^2) \delta (\partial_th_3+j^3),
\end{equation}
which can be used to integrate over $h_2$ and $h_3$.
\begin{equation}
Z(J,j)=(\det \partial_t)^{-2}\exp i\int dt (-J^2{\partial_t}^{-1}j^2
-J^3{\partial_t}^{-1}j^3-\frac {i\Gamma_2}2 j^2
-\frac {i\Gamma_3}2 j^3 ) \plabel{Z2}
\end{equation}
where in $\Gamma_2$ and $\Gamma_3$ the triad fluctuations are expressed in
terms of the currents, $h_{2,3}=-\partial_t^{-1}j^{2,3}$.

Let us now calculate the effective action
\begin{equation}
W_{eff}(\bar h,\bar \phi )=\frac 1i {\rm ln} Z-
\bar h_2J^2-\bar h_3J^3-\bar \phi_2j^2-\bar \phi_3j^3.
\plabel{Weff}
\end{equation}
The currents should be expressed in terms of mean field by the
equations
\begin{eqnarray}
\bar h_2 &=&\frac{\delta}{i\delta J^2} {\rm ln}Z=
-\partial_t^{-1}j^2
\nonumber \\
\bar h_3 &=&\frac{\delta}{i\delta J^3} {\rm ln}Z=
-\partial_t^{-1}j^3 \plabel{cur}
\end{eqnarray}
The field independent infinite factor $(\det \partial_t)^{-2}$ does
not contribute to the effective action. By substituting (\ref{Z2})
and (\ref{cur}) in (\ref{Weff}) we obtain
\begin{equation}
W_{eff}(\bar h,\bar \phi )=\int dt (-\bar h_2 \partial_t\bar \phi_2 -
\bar h_3 \partial_t\bar \phi_3 +\frac i2 \Gamma_2\partial_t\bar h_2+
\frac i2 \Gamma_3\partial_t\bar h_3)
\plabel{Weff1}
\end{equation}
This means that the full effective action (\ref{Weff1})
coincides with the classical action on the reduced phase
space.

Let us consider now the case of triad fluctuations with
coinciding eigenvalues. The Faddeev--Popov determinant is
\begin{equation}
\det \{ G_A,\chi^B\} = -2(h_2\phi_2+h_3\phi_3)h_1
(h_1^2-h_2^2)(h_3^2-h_2^2)
(h_1^2-h_3^2). \plabel{FaPo}
\end{equation}
According to our gauge condition $h_1=1$ but we prefer to
keep more symmetric notations in (\ref{FaPo}). In the case of
the increased symmetry at least
two eigenvalues among $h_1,\ h_2,\ h_3$ coincide, and the
determinant (\ref{FaPo}) is zero. This effect was noted
by Mottola \cite{Mot} in a different context. He claimed
that the path integral measure is zero on symmetric
configurations and thus they do not contribute to the path
integral. We see that only the first part of this conjecture
is true. The  zero in (\ref{FaPo}) is cancelled 
by the contribution of the delta function of the constraints. Hence the path
integral measure is regular near symmetric configurations.

Note, that singularities in the Faddeev--Popov determinant
can disappear if one uses the gauge fixing approach instead of
the reduced phase space one. For example, if one replaces
the condition (\ref{hdiag}) by $\tilde \eta^a=n^a=0$ the
Faddeev--Popov determinant (calculated e.g. in the BFV
approach \cite{He}) will contain $(\det \partial_0)^6$.
This multiplier does not depend on a symmetry of the three space
geometry.

\section{Conclusions}
In the present paper we suggested an $SU(2)$ invariant
reduction of the $3+1$-dimensional Ashtekar gravity. We
assumed that spatial sections are isomorphic to $S^3$,
which is the group manifold of $SU(2)$. The invariant fields
have constant components in the canonical tagential basis.
We constructed a reduced phase space quantization of
this model. We observe that near the triad configurations
admitting additional Killing vectors the Faddeev--Popov
determinant is  zero. This  zero is cancelled by the
contribution of the delta functions of the constraints. 
The resulting path integral measure is regular near 
symmetric points in the reduced phase space. Moreover, 
in our simple model we are able to calculate generating 
functional of Green functions of the reduced phase space 
variables. It occurs that at all orders of the perturbation 
theory this functional contains tree--level diagrams only.
The corresponding quantum effective action coincides with 
the classical action calculated on the
reduced phase space.

\section{Acknowledgments}
This work was partially supported by GRACENAS and
Russian Foundation for Fundamental Research,
grant 97-01-01186.

\end {document}